# Path Following Control of Automated Vehicle Considering Model Uncertainties External Disturbances and Parametric Varying

*Dan Shen, Electrical and Computer Engineering at Purdue University*

**Abstract:** Automated Vehicle Path Following Control (PFC) is an advanced control system that can regulate the vehicle into a collision-free region in the presence of other objects on the road. Common collision avoidance functions, such as forward collision warning and automatic emergency braking, have recently been developed and equipped on production vehicles. However, it is impossible to develop a perfectly precise vehicle model when the vehicle is driving. Most PFCs did not consider uncertainties in the vehicle model, external disturbances, and parameter variations at the same time. To address the issues associated with this important feature and function in autonomous driving, a new vehicle PFC is proposed using a robust model predictive control (MPC) design technique based on matrix inequality and the theoretical approach of the hybrid $\&$ switched system. The proposed methodology requires a combination of continuous and discrete states, e.g. regulating the continuous states of the AV (e.g., velocity and yaw angle) and discrete switching of the control strategy that affects the dynamic behaviors of the AV under different driving speeds. Firstly, considering bounded model uncertainties, and norm-bounded external disturbances, the system states and control matrices are modified. In addition, the vehicle's time-varying longitudinal speed is considered, and a vehicle lateral dynamic model based on Linear Parameter Varying (LPV) is established by utilizing a polytope with finite vertices. Then the Min-Max robust MPC state feedback control law is obtained at every timestamp by solving a set of matrix inequalities which are derived from Lyapunov stability and the minimization of the worst-case in infinite-horizon quadratic objective function. Compared to adaptive MPC, nonlinear MPC, and cascade LPV control, the proposed robust LPV MPC shows improved tracing accuracy along vehicle lateral dynamics. Finally, the state feedback switched LPV control theory with separate Lyapunov functions under both arbitrary switching and average-dwell-time (ADT) switching conditions are studied and applied to cover the path following control in full speed range. Numerical examples, tracking effectiveness, and convergence analysis are provided to demonstrate and ensure the control effectiveness and strong robustness of the proposed algorithms.

# Introduction

## 1.1 Background and Motivation

Based on recent statements by the World Health Organization (WHO), about 1.25 million people lose their lives every year due to traffic accidents [1]. In recent years, AVs have been developed rapidly and become an attractive research area in both the automotive industry and academia with the objective of ensuring driving safety and improving passenger comfort. With other major breakthroughs in science and technology such as sensing technology and powerful computing capabilities, as well as the rapid development of data sources, decision-making and intelligent control, and human-machine interface, AVs can have a much broader spectrum of applications in future Intelligent Transportation Systems (ITS). Thus, AV technology has been widely regarded as a practical solution of potentially mitigating human driving errors and reducing the number of accidents on roads [2], [3].

Early research on AVs focused on developing Advanced Driver Assistance Systems (ADAS) to help drivers with decision-making and control to improve driving safety. The latest state-of-the-art ADAS technologies, e.g., Autonomous Emergency Brake System (AEBS) and Lane centering system (LCS), have made it possible to produce reliable self-driving vehicles, that can free the human driver from the heavy driving workload and reduce vehicle crashes. The National Highway Traffic Safety Administration (NHTSA) reported that rear-end crashes are one of the common types of traffic incidents on U.S. highways, accounting for approximately 29% of all crashes and resulting in a substantial number of injuries and fatalities every year [3]. Recent studies have found that there are many factors that can contribute to these crashes, such as low visibility, slippery road surfaces, driver distraction, drowsy driving, and so on [4]–[6]. Meanwhile, crashes due to road departure is a leading cause of fatalities on roads in the U.S., which accounts for about 35% of the fatalities [7], [8]. To alleviate both casualties and economic losses due to accidents, many ADAS functions have been developed and equipped on production vehicles, such as road departure mitigation system [2], and electric stability control (ESC) system [9].

To accomplish the goal of fully autonomous driving, a set of necessary subsystems need to work collaboratively. Firstly, the perception system (e.g., cameras, radar, LiDAR, etc.) is used to obtain environmental information, such as the locations of obstacles, distances to surrounding vehicles, existence of pedestrians, etc. Then, the risk assessment module computes the danger of potential paths of the ego vehicle based on the environmental information, and the motion planning module determines the most desirable vehicle trajectory that meets both safety and comfort requirements. The desired trajectory consists of the positions, velocity, and orientation of the vehicle in time series [10]. Finally, the control system takes over and generates the constrained control actions (acceleration, braking, and steering wheel) for the actuators to allow the vehicle following the pre-defined trajectory. Therefore, the control system is a very crucial module of AVs. Although control theory is relatively mature and has been applied in many fields, the control strategies of AVs still need to be further studied because of many existing challenges and uncertainties.

From the model-based control perspective, the vehicle control problem is mainly related to two features: the control categories (longitudinal control, lateral control, and integrated control) and the model considered in the controller design (kinematic model or dynamic model). This work focuses on the coupled longitudinal and lateral control problem using both vehicle kinematic and dynamic models. The longitudinal control of an AV aims to track the vehicle's desired trajectory automatically and robustly through the throttle and braking control by considering varying velocity and road conditions. Indeed, longitudinal control plays a significant role in ensuring the safety and comfort of passengers [11], where a nonlinear longitudinal control design based on a Lyapunov approach is proposed by taking into account the throttle, brake, and gear ratio. In [12], the longitudinal velocity control problem for AVs is addressed using deterministic promotion reinforcement learning, which contains car-following and non-car-following conditions in a unitized form.

The lateral control of an AV aims to make the ego vehicle follow the predetermined paths automatically and ensure stability and robustness against various driving conditions [13]. Several researchers studied the active front steering (AFS) system, which utilizes the vehicle front steering command to enhance its stability and compensate the appropriate amount of angle to the drivers' steering wheel angle in some extreme conditions [14]. In [15], the authors investigated a path-following application for vehicles based on a simple linear and time-invariant single-track model, which is calculated based on a constant nominal longitudinal speed. The authors of [16] proposed a nonlinear model predictive control (NMPC) method that integrates AFS and additional yaw movement to obtain active chassis lateral stability. The aforementioned works have also made great progress in the stability analysis of vehicle lateral control, such as the shift-able stability regions and dynamic margins [17].

Integrated control of AVs is a combined control scheme with both longitudinal and lateral control. The longitudinal controller is mainly responsible for regulating the vehicle's cruise velocity, while the lateral controller steers the vehicle's wheels for path tracking [18]. The nonlinear coupling relationship between the vehicle's longitudinal and lateral dynamics imposes a significant challenge on the implementation of the coupled control. Recently, many researchers have been working on this topic. In [19], the authors developed an integrated longitudinal and lateral trajectory planning and tracking control algorithm under vehicle-to-vehicle communication. The problem of coordinated lateral and longitudinal vehicle-following control for connected and automated vehicles (CAVs) was investigated by considering nonlinear dynamics in [20]. As mentioned above, the accurate trajectory tracking control in AVs requires both the steering wheel control for regulating the lateral position and the throttle/braking control for adjusting its longitudinal dynamics. A vehicle trajectory tracking method with a time-varying model based on the model predictive control using the vehicle kinematic model was proposed in [21]. Random network delay was introduced in [22] to present the uncertainty of the trajectory tracking model of the AV, which significantly deteriorates the stability of the control system and accuracy of the trajectory tracking. In [23], the authors proposed a human-centered trajectory tracking control strategy that integrates driver behavior prediction for the cut-in scenarios and their transient processes.

While in real applications, it is impossible to establish an accurate model when the vehicle is moving in an unpredictable environment. Besides, the time-varying model uncertainties such as nonlinear characteristics of tire model, external disturbances of wind gusts or road gradient, and parametric varying of vehicle velocity will seriously affect the performance of vehicle dynamic control. Therefore, more robust control approaches have been investigated by researchers. As one of the potential solutions, the control theory using Linear Parameter Varying (LPV) representations aims to solve nonlinear control problems using a pseudolinear reformulation that incorporates the original nonlinearity within certain scheduling parameters [24]. Several control design approaches can be used such as pole positioning, H-infinity, H2, and LQR. The authors of [25], and [26] have contributed significantly to LPV modeling and H-infinity control. The authors in [1] proposed a vehicle automated steering controller based on the MPC approach by considering the parametric uncertainties of velocity varying. To deal with the presence of steering actuation nonlinearity, the authors in [27] developed an LPV MISO H-infinity controller based on the feedback of the lateral error at the center of gravity and the look-ahead distance. Except for the LPV models, the Takagi- Sugeno (T-S) model-based MPC control utilizing LMI was also developed in [28], [29]. As proposed in [30], a general T-S vehicle lateral dynamic control framework was designed with considering the norm-bounded uncertainty from parametric varying by using the variations of vehicle mass, longitudinal speed, nonlinearity of tire, and the road forces.

Besides the above-mentioned LPV techniques, robust model predictive control has also been developed and explored to deal with the model uncertainties or external disturbances in vehicle control. A robust feedback MPC was implemented on an elastic flight vehicle in [31]. In [32], the adaptive tube MPC has been applied for efficient cruise control on a hybrid plug-in electric vehicle. The authors in [33] proposed a min-max robust MPC method for vehicle trajectory tracking in the presence of slip angle. Similarly, Liu in [34] designed a feedback min-max MPC for semi-autonomous vehicle lane keeping. The most recent work in [35] has introduced an LMI and BMI-based robust MPC to handle the vehicle control with model uncertainties. As can be seen in the literature, since the optimization problems using LMI or BMI are normally convex and their global solvable solutions can be computed efficiently and applicably by using some online solvers, the LMI or BMI-based robust MPC techniques have been applied widely in the control of autonomous vehicles. However, most of the current existing work applying the LMIs or BMIs-based robust MPC only focuses on the vehicle path following control considering the model uncertainties. Few papers on matrix inequalities based robust MPC take both the model uncertainties and external disturbances of the vehicle into account. The research in [1] applied the LMI-based robust MPC for vehicle trajectory tracking considering both model uncertainties of vehicle tire stiffness and the parametric varying of longitudinal velocity. However, the external disturbance was not included in the control design. Another paper [36] has made further contributions and explored the matrix inequalities based robust MPC for vehicle lateral dynamics tracking considering model uncertainties, external disturbances, and time-varying delay. However, the parametric varying issue still remains by making the assumption of constant vehicle speed during the control process.

Moreover, it is noted in the literature, most of the path tracking control problems assume a constant vehicle speed and only the vehicle lateral dynamic is considered to obtain the desired control actions of the steering wheel angle. Some previous works explored the coupled control with a simplified vehicle model in both longitudinal and lateral directions. For instance, [25], [29] utilized the cascade control configuration for vehicle trajectory following based on TS model and LPV model with consideration of velocity-varying. However, the model uncertainties and external disturbances were not considered in the vehicle model and motion control design. Moreover, the vehicle lateral dynamics will be affected significantly under varying longitudinal speed. However, the most above-mentioned path-following controller only has one set of parameters for trajectory tracking with large longitudinal speed variations in the model, which will degrade the control performance. Thus, the parameter-dependent switching control strategy may be a better choice to handle the parameter variations.

Although the aforementioned works contributed significantly in this research direction, the automated vehicle trajectory tracking control still has many issues that remain to be resolved [37], [38]. Thus, the main objective of this dissertation is to develop a matrix inequalities based robust MPC control for vehicle trajectory tracking by considering bounded model uncertainties, norm-bounded external disturbances, and the bounded parametric varying in terms of vehicle longitudinal velocity, which can ensure the accuracy of path-tracking and its robustness on both straight and curved roads. Both the longitudinal control and lateral control will be involved in the path following. Two other state-of-the-art methods of vehicle trajectory following control will also be introduced and applied based on the previous experiences as comparisons with the proposed technique, which are adaptive MPC and cascade LPV control [39]. Meanwhile, the state feedback switching LPV control theory will also be studied and applied to cover the path following control in full speed range using the vehicle lateral error dynamic model [40]. All the controller synthesis conditions are presented and expressed with parameter-dependent linear matrix inequalities.

*This is the partial of Dr. Dan Shen's Ph.D. dissertation. The full version will be released in April 2024.*